# Multiple Imputation for Small, Extremely High Efficacy Clinical Trials with Binary Endpoints

Yaoyuan Vincent Tan*, Gang Xu, Chenkun Wang

*corresponding author

Affiliation: Vertex Pharmaceuticals

## 1. Introduction

Despite the sponsor's best efforts, it is not uncommon to see missing data in clinical trials. When data is Missing Completely At Random (MCAR), or when the amount of missing data is a small portion of the sample size, the impact of missing data will be minimal. But when this is otherwise, missing data will bias analysis and inference. In 2012, the Food and Drug Administration (FDA) provided an important and influential guidance (Little et. al., 2012) regarding the handling of missing data, emphasizing that prevention is better than solution. In the event missing data still occurs, a commonly recommended method to handle it is multiple imputation (MI; Rubin, 1987). MI is often preferred over single imputation methods for example, last observation carried forward, because MI is able to adequately capture the uncertainty from missing data. Recently, there has been growing interest in the use of Cell and Gene Therapy (CGT) to develop treatments. The hallmark of a CGT clinical trial is the extremely high efficacy demonstrated (Frangoul et al., 2024; Locatelli et al., 2024, Hering et al., 2016). Unfortunately, such trials are expensive to conduct, indicated for rare diseases, with the expectation of transformative effect, and hence, have extremely small sample sizes and often, a single arm design. Complicating this situation is the presence of missing data coupled with high unmet need. Due to the novelty of these therapies, regulators will often request sponsors to handle missing data in the analysis even if the amount of missing data is small. Unsurprisingly, the common method used to account for missing data is MI.

MI for small sample size is not an uncommon problem. Barnard and Rubin (1999) suggested an adjustment to the *t*-distribution used to approximate MI to mitigate this problem. Barnes, Lindborg, and Seaman (2006) also discussed and compared various techniques for implementing MI on small sample clinical trials. MI literature for continuous endpoints is well established. This is unsurprising as MI was developed by using the normal distribution (*t*-distribution for more robustness) to approximate the posterior distribution. Literature for implementing MI on binary endpoints is not as common in contrast. One straight forward way to handle binary endpoints is to log-odds transform the binary endpoint such that standard MI methods can be applied. Lott and Reiter (2020) proposed a Wilson interval (Wilson, 1927) for MI by directly using the mean and variance of the binomial distribution to calculate the conditional mean and variance for the *t*-distribution used to approximate MI. Mandan (2018) proposed Jeffreys (Jeffreys, 1946) and Clopper-Pearson (CP; Clopper and Pearson, 1934) intervals by using Zhou and Reiter (2010)'s recommendation for Bayesian inference on MI for the Jeffreys interval and fixing the observed success as the mean success, mean limit, and mean cumulative probability in order to calculate the CP interval. Although there are existing literatures to solve the issue of implementing MI for binary endpoints under small sample clinical trial conditions, literature discussing implementation of MI for studies with extremely high success rate is scarce.

In this work, motived by the characteristics of CGT trials, we are interested in implementing MI for small sample clinical trials with binary endpoints and extremely high success rates. Our parameter of interest is the response rate often estimated as the number of successes divided by the sample size. We investigated several existing methods for implementing MI for binary endpoints and proposed three additional methods. The first method is an algorithmic approach by modifying MI using Clopper-Pearson intervals. This method ensures that the produced 95% CI will be conservative. The second is a fully Bayesian approach. Our third approach is based on the observation that standard MI approximates the posterior distribution using a normal distribution. We instead propose a beta distribution for the approximation. We briefly review MI in the next section followed by a detailed exposition of our three proposed methods. In Section 3, we compare several missing data handling methods including our three proposed methods under small sample size and extremely high success rate scenarios using simulations to measure and quantify their performance. In Section 4, we apply our three proposed methods together with complete case analysis, single imputation as all success, single imputation as all failure, and MI using Wald interval, to the primary endpoint of the Clinical Islet Transplantation (CIT) Protocol 07 (Hering et al., 2016). Finally, we provide some insights and possible future work in Section 5.

## 2. Methods

### 2.1 Multiple imputation

Readers who are interested in a comprehensive exposition on MI can refer to Rubin, 1987; Little and Rubin, 2019; and Murray, 2018. Here, we provide a brief summary. The theory and motivation for MI is based on the Bayesian posterior distribution for the parameter of interest. Let $\pi$ be the parameter of interest, $Y^{obs}$ denote the observed data in the study and $Y^{mis}$ denote the missing data in the study. Then without loss of generality, given the missingness mechanism is Missing At Random (MAR), the posterior distribution of $\pi$ in the presence of missing data can be written as

$$p(\pi|Y^{obs}) = \int_{Y^{miss}} p(\pi|Y^{obs}, Y^{miss}) p(Y^{miss}|Y^{obs}) dY^{miss}. \tag{1}$$

Equation (1) is the heart and core of MI. There are two important components of Equation (1). First $p(Y^{miss}|Y^{obs})$, the posterior predictive distribution and second, $p(\pi|Y^{obs}, Y^{miss})$, the posterior draw of $\pi$ given the observed and missing data. This implies that to obtain a draw from the posterior distribution of $p(\pi|Y^{obs})$ in the presence of missing data, we must first impute the missing data, $Y^{mis}$, using the posterior predictive distribution. Once $Y^{mis}$ is obtained, we can estimate $\pi$ by using the observed data $Y^{obs}$ and the imputed data $Y^{mis}$. An important note here is

that $p(Y^{miss}|Y^{obs})$ and $p(\pi|Y^{obs}, Y^{miss})$ are draws from the posterior distribution and posterior predictive distribution respectively. Hence, substituting them by using estimates from other modeling or optimization techniques like regression models may not constitute a valid MI (Little and Rubin, 2019 Section 10; Murray, 2018). However, when the amount of missing data is small, such strategies would not affect the results and conclusions of MI (Little and Rubin, 2019 Section 10). In limited cases, the integration of Equation (1) will be tractable and instead of implementing two major draws, one single draw from $p(\pi|Y^{obs})$ suffices. Similarly, in limited cases, $p(\pi|Y^{obs}, Y^{miss})$ and $p(Y^{miss}|Y^{obs})$ may have closed form solutions. When closed form solutions are not available, $p(\pi|Y^{obs}, Y^{miss})$ may be expanded and written as

$$p(\pi|Y^{obs}, Y^{miss}) = \frac{p(Y^{obs}, Y^{miss}|\pi)p(\pi)}{p(Y^{obs}, Y^{miss})} \quad (2)$$

and $p(Y^{miss}|Y^{obs})$ can be expanded to become

$$p(Y^{miss}|Y^{obs}) = \int_{\pi} p(Y^{miss}|\pi) \frac{p(Y^{obs}|\pi)p(\pi)}{p(Y^{obs})} d\pi \quad (3)$$

where drawing from Equations (2) and (3) can be handled using Gibbs sampling or Metropolis-Hastings (MH) methods where applicable.

Drawing from Equation (1) typically requires at least 10,000 draws to ensure the posterior distribution is sufficiently profiled. From a practical standpoint, this may not be ideal. The ingenuity of MI is to recognize that often, the eventual distribution of $p(\pi|Y^{obs})$ will be a normal distribution or can be approximated well using a normal distribution. Thus, a good estimation of $p(\pi|Y^{obs})$ can be obtained using much lesser number of draws, $D$, say 5 or 10. In addition, since Equation (1) can be approximated well using a normal distribution, only the sufficient statistics, the mean and variance, are needed in order to fully describe $p(\pi|Y^{obs})$. The mean and variance of $p(\pi|Y^{obs})$ can be calculated using

$$E[\pi|Y^{obs}] = E\big[E(\pi|Y^{obs}, Y^{miss})\big|Y^{obs}\big] \quad (4)$$

and

$$Var[\pi|Y^{obs}] = E\big[Var(\pi|Y^{obs}, Y^{miss})\big|Y^{obs}\big] + Var\big[E(\pi|Y^{obs}, Y^{miss})\big|Y^{obs}\big] \quad (5)$$

respectively. Hence, using $D$ multiply imputed datasets of $\pi$, the famous Rubin's combine rules can be obtained

$$E[\pi|Y^{obs}] \approx \int \pi \frac{1}{D} \sum_{d=1}^{D} p(\pi|Y^{obs}, Y^{miss,(d)}) d\pi = \bar{\pi};$$

$$Var[\pi|Y^{obs}] \approx \frac{1}{D} \sum_{d=1}^{D} V_d + \frac{1}{D-1} \sum_{d=1}^{D} (\widehat{\pi^d} - \bar{\pi})^2 = \bar{V} + B \qquad (6)$$

Where $\widehat{\pi^d}$ is the posterior mean of $\pi$ calculated for the $d^{th}$ dataset, $V_d$ is the complete-data posterior variance of $\pi$ calculated for the $d^{th}$ dataset and $\bar{V}$ is the average of $V_d$ over the $D$ datasets and $B$ is the between-imputation variance. The 95% confidence interval (CI) can then be calculated using the estimated posterior mean and variance from Equation (6) together with the normal distribution assumption. Various adjustments to Equation (6) can be made to enhance the performance when $D$ is small or when the amount of missing data is small (Details refer to Little and Rubin, 2019 Section 5).

2.2 Modified Clopper-Pearson Multiple Imputation intervals

Our proposed modified Clopper-Pearson Multiple Imputation (CPMI) method is algorithmic in nature aimed at producing a conservative 95% CI. The motivation behind this method is the observation that CP CIs obtained from a single imputation are even wider than MI CIs. For example, suppose we conduct a single arm trial with a binary endpoint and the parameter of interest is the response rate. We recruited 20 subjects, 12 responded with success, 7 failed to respond, but the endpoint for 1 subject was missing. If we imputed this missing data as a success, the estimated success rate would be 65% and the 95% CP CI is (41%, 85%) with a CI length of 44 percentage points (pp). Imputing as a failure, the estimated success rate is 60% with a 95% CP CI of (36%, 81%), anda CI length of 45 pp. Using standard MI, with 20 multiply imputes ($D = 20$), the resulting estimate is 64% with a 95% CI of (42%, 85%), with a CI length of 43 pp. This is counter intuitive as we expect MI to capture uncertainty from missingness which should be producing wider CI compared to single imputation methods. The driving reason for this phenomenon is that CP CIs are recommended for small sample size trials and are overly conservative due to the discrete nature of binomial distribution. Unlike continuous distributions like normal distribution where the upper and lower CI limits can be adjusted smoothly such that crossing of the 5% threshold can be achieved without huge jumps in probability, for the binomial distribution, the crossing of the 5% threshold for the 95% CI comes in steps and the smaller the sample size, the larger the step. Moreover, MI relies on the normal approximation. Hence, the larger the sample size, the better the approximation. The sample size of 20 or less does not help.

Our conservative CPMI algorithm is constructed as follows:

i. Draw and impute the missing data using standard MI procedures with the posterior predictive distribution.
ii. For the imputed dataset, compute the success rate estimate for the $d^{th}$ dataset, $\pi^d$, and compute the CP CI, denoting the lower and upper bound as $\pi_{lb}^d$ and $\pi_{ub}^d$ respectively.

iii. Using the success rate from the imputed dataset in Step ii, draw $n$ responses where $n$ is the sample size.

iv. Calculate the corresponding CP CI $(\pi_{lb*}^{dd}, \pi_{ub*}^{dd})$ from the $n$ responses in Step iii.

v. Repeat Steps iii and iv $DD$ times to obtain $Var(\pi_{lb}^d) = \frac{1}{DD-1}\sum_{d=1}^{DD}(\pi_{lb*}^{dd} - \bar{\pi}_{lb*})^2$ and $Var(\pi_{ub}^d) = \frac{1}{DD-1}\sum_{d=1}^{DD}(\pi_{ub*}^{dd} - \bar{\pi}_{ub*})^2$ where $\bar{\pi}_{lb*} = \frac{1}{DD}\sum_{d=1}^{DD}\pi_{lb*}^{dd}$ and $\bar{\pi}_{ub*} = \frac{1}{DD}\sum_{d=1}^{DD}\pi_{ub*}^{dd}$.

vi. Repeat Steps i to v to obtain $D$ multiply impute datasets.

vii. Use Equation (6) with correction for small $D$ to calculate the total variation $T_{lb}^D$ and $T_{ub}^D$ by using $V_d$ as $Var(\pi_{lb}^d)$ or $Var(\pi_{ub}^d)$ in Equation (6) and $B$ using $\pi_{lb}^d$ or $\pi_{ub}^d$, respectively.

viii. Compute $\bar{\pi} = \frac{1}{D}\sum_{d=1}^{D}\pi^d$ as well as $\bar{\pi}_{lb} = \frac{1}{D}\sum_{d=1}^{D}\pi_{lb}^d$ and $\bar{\pi}_{ub} = \frac{1}{D}\sum_{d=1}^{D}\pi_{ub}^d$.

ix. The modified CPMI is $(\bar{\pi} - \bar{\pi}_{lb} - t_\nu\sqrt{T_{lb}^D}, \bar{\pi} + \bar{\pi}_{ub} + t_\nu\sqrt{T_{ub}^D})$.

Implementing the modified CPMI on our example, we have an estimated success rate of 63% with a 95% CI of (37%, 89%), and a CI length of 52 pp.

## 2.3 Fully Bayesian approach for Binary Endpoints

A fully Bayesian approach for handling imputation of missing binary endpoints in the inference of success rates is nothing but drawing from Equation (1). More specifically, we need to fully specify the distributions of the parameters and random variables in Equation (1). Let $p(\pi)$ be the prior distribution of the success rate pre-imputation, $p(\pi')$ be the prior distribution of the success rate post-imputation, $n$ denote the sample size, $n^{obs}$ denote the number of subjects with complete data, and $n^{miss}$ denote the number of subjects with missing data. Then a fully Bayesian parametrization will be

$$Y^{obs}, Y^{miss}|\pi' \sim Bin(n, \pi'),$$
$$\pi' \sim Beta(\alpha', \beta'),$$
$$Y^{obs}|\pi \sim Bin(n^{obs}, \pi),$$
$$Y^{miss}|\pi \sim Bin(n^{miss}, \pi), \text{ and}$$
$$\pi \sim Beta(\alpha, \beta).$$

This gives

$$Y^{miss}|Y^{obs} \sim Beta-Binomial(n^{miss}, \alpha + y^{obs}, \beta + n^{obs} - y^{obs}) \quad (7)$$
$$\pi|Y^{obs}, Y^{miss} \sim Beta(\alpha' + y^{obs} + y^{miss}, \beta' + n - y^{obs} - y^{miss})$$

where $y^{obs}$ is the observed number of success and $y^{miss}$ is the imputed number of success drawn from the posterior predictive distribution. Drawing from $Y^{miss}|Y^{obs}$ followed by $\pi|Y^{obs}, Y^{miss}$, $M$ times will allow us to characterize the profile of the posterior distribution $p(\pi|Y^{obs})$. An empirical CI can then be obtained by $I_{emp}(\pi) = (\hat{\pi}^{(m,\frac{\alpha}{2})}, \hat{\pi}^{(m,1-\frac{\alpha}{2})})$, where $\hat{\pi}^{(m,\frac{\alpha}{2})}$ and $\hat{\pi}^{(m,1-\frac{\alpha}{2})}$ are the empirical $m^{th}$ draw from the posterior distribution $p(\pi|Y^{obs})$. Using the fully Bayesian approach described above on our simple example, with $M = 100{,}000$, we obtained an estimate of 63% with a CI of (41%, 82%), and a CI length of 41 pp.

2.4 Beta approximation Multiple Imputation

Although a fully Bayesian approach has many desirable qualities (nominal 95% coverage and shortest 95% CI length [as our simulation results show]), the prohibitive number of draws and hence datasets that a sponsor will have to perform and store may prove to be a price that is too high to pay. Hence, an alternative MI-like procedure was sought after. Following a similar trend of thought as standard MI, we write out the parametrization in Section 2.3 to obtain the following fully parameterized posterior distribution $p(\pi|Y^{obs})$ as

$$p(\pi|Y^{obs}) = \int_{Y^{miss}} p(\pi|Y^{obs}, Y^{miss}) p(Y^{miss}|Y^{obs}) dY^{miss} \qquad (8)$$

$$= \int_{Y^{miss}} \frac{\Gamma(\alpha' + \beta' + n)}{\Gamma(\alpha' + Y^{obs} + Y^{miss})\Gamma(\beta' + n - Y^{obs} - Y^{miss})} \pi^{\alpha' + Y^{obs} + Y^{miss} - 1} (1-\pi)^{\beta' + n - Y^{obs} - Y^{miss} - 1}$$
$$\times \frac{\Gamma(n^{miss} + 1)}{\Gamma(Y^{miss} + 1)\Gamma(n^{miss} - Y^{miss} + 1)} \frac{\Gamma(\alpha + Y^{obs} + Y^{miss})\Gamma(n + \beta - Y^{obs} - Y^{miss})}{\Gamma(\alpha + \beta + n)}$$
$$\times \frac{\Gamma(\alpha + \beta + n^{obs})}{\Gamma(\alpha + Y^{obs})\Gamma(\beta + n^{obs} - Y^{obs})} dY^{miss}.$$

From Equation (8), we can see that $p(\pi|Y^{obs}) \propto \pi^{\alpha' + Y^{obs} + Y^{miss} - 1}(1-\pi)^{\beta' + n - Y^{obs} - Y^{miss} - 1}$ which is in the form of a Beta distribution where $Y^{miss}$ will need to be integrated out. As such, we make the bold assumption that Equation (8) can be well approximated by a Beta distribution, similar to the strategy adopted by (Rubin, 1987) in his development of MI using the normal distribution approximation. Along the same vein, we will use Equation (6) to obtain an estimate for the posterior mean and variance and use these estimates to calculate the parameters of our Beta distribution approximation. Notate the posterior mean estimate as $\hat{\mu}^*$ and posterior variance estimate as $\widehat{\sigma^2}^*$, the estimated parameters for the Beta distribution that we use to approximate posterior distribution $p(\pi|Y^{obs})$ is then $\hat{\alpha}^* = \hat{\mu}^*(\frac{\hat{\mu}^*(1-\hat{\mu}^*)}{\widehat{\sigma^2}^*} - 1)$ and $\hat{\beta}^* = (1 - \hat{\mu}^*)(\frac{\hat{\mu}^*(1-\hat{\mu}^*)}{\widehat{\sigma^2}^*} - 1)$. The 95% CI can then be obtained using a grid search of this Beta distribution to find a CI with 95% coverage and the shortest length. In practice, there could be a few such intervals with the shortest length identified. To be conservative, the interval with the smallest 95% CI lower bound value can be chosen (assuming higher success rates indicate better prospects for the therapy). With this approach, the estimated success rate for our simple example is 61% and the 95% CI is (28%, 92%), with a CI length of 64 pp.

# 3. Simulation

3.1 Quantifying the performance of our proposed methods and various existing methods using simulation

We ran extensive simulations to determine the performance of various missing data and MI methods together with our three proposals under small sample sizes and extremely high success rates. We compared

  a) Complete case analysis,
  b) Impute as success – single imputation,
  c) Impute as failure – single imputation,
  d) Multiple imputation using Wald intervals,
  e) Multiple imputation using Wilson intervals,
  f) Multiple imputation using log-odds transformation with Barnard and Rubin (1999) adjustment,
  g) Bootstrap (where parameter of interest is the arithmetic success rate),
  h) Jackknife (where parameter of interest is the arithmetic success rate),
  i) Modified Clopper-Pearson multiple imputation (Modified CPMI),
  j) Full Bayesian imputation, and
  k) Multiple imputation using Beta distribution.

Specific details of how each method was implemented, and the simulation codes are provided in the Appendix and supplementary materials respectively. The missingness mechanism we employed is Missing Completely At Random (MCAR). This is because MAR and Missing Not At Random (MNAR) assumptions largely impact the bias rather than the method's ability to account for missingness uncertainty. Identification and verification of the missingness mechanism is a separate topic beyond the scope of this work. Hence, we argue that considering MCAR scenarios will suffice to help us understand the impact of these methods for capturing the missingness uncertainty. The statistics of interest are the average 95% CI length and 95% coverage produced by each method over all simulations for each scenario. A total of 64 simulation scenarios were investigated by varying the missing rates (1%, 10%, 20%, 30%), sample sizes (10, 20, 30, 50), and true success rates (70%, 80%, 90%, 99%).

3.2 Results

Table 1 summarizes the results over all the simulation scenarios we investigated. Not surprisingly, the fully Bayesian method performed the best with one of the shortest average 95% CI length (mean of 28 pp) and a nominal coverage probability close to 95% (mean and median) for most of the simulation scenarios. MI using Beta distribution approximation, modified CPMI, complete case analysis, and MI using Wilson intervals also produced satisfactory results. Complete case analysis, modified CPMI, and MI using Beta distribution approximation were more conservative, with minimum coverage probability of 95% or close to 95%. Although complete case analysis produced adequate results which were conservative, this is not surprising as we assumed MCAR in our simulations. In practice, if researchers would like to use complete case analysis, the MCAR assumption should be verified as violation of this assumption will lead to bias. MI using Wilson intervals although satisfactory, will produce 95% CI that are below coverage when the true success rate is extremely high, 99%, indicating the breakdown of the approximations used when constructing the Wilson MI intervals. Finally, 95% coverage for the rest of the missing data handling methods were too low, between 0 to 75% indicating severe under-coverage in some of our simulation scenarios.

Table 1: Summary statistics (mean, median, minimum, and maximum) of average 95% confidence interval (CI) length and 95% CI coverage over all simulation scenarios investigated.

| Methods | Average 95% CI length | | | | 95% CI coverage probability (%) | | | |
|---|---|---|---|---|---|---|---|---|
| | Mean | Median | Minimum | Maximum | Mean | Median | Minimum | Maximum |
| a) Complete case | 0.34 | 0.33 | 0.09 | 0.65 | 98 | 98 | 96 | 100 |
| b) Impute as success | 0.29 | 0.28 | 0.08 | 0.55 | 97 | 99 | 75 | 100 |
| c) Impute as failure | 0.38 | 0.36 | 0.12 | 0.6 | 59 | 74 | 0 | 98 |
| d) Multiple imputation using Wald intervals | 0.28 | 0.27 | 0.04 | 0.62 | 89 | 92 | 47 | 100 |
| e) Multiple imputation using Wilson intervals | 0.33 | 0.31 | 0.1 | 0.55 | 95 | 96 | 87 | 98 |
| f) Multiple imputation using log-odds transformation | 0.41 | 0.36 | 0.14 | 0.74 | 94 | 97 | 35 | 100 |
| g) Bootstrap | 0.26 | 0.26 | 0.03 | 0.63 | 72 | 93 | 7 | 97 |
| h) Jackknife | 0.49 | 0.5 | 0.05 | 1.02 | 73 | 90 | 7 | 100 |
| i) Modified Clopper-Pearson multiple imputation | 0.4 | 0.39 | 0.11 | 0.77 | 99 | 99 | 95 | 100 |
| j) Fully Bayesian approach | 0.28 | 0.27 | 0.07 | 0.55 | 95 | 95 | 92 | 99 |
| i) Multiple imputation using Beta distribution | 0.34 | 0.34 | 0.07 | 0.66 | 98 | 98 | 94 | 100 |

To understand how these methods perform as the missing rate, sample size and true success rate vary, we created barplots to help visualize the results. Figure 1 shows the average 95% CI length and 95% coverage when the true success rate is varied (10% missing rate, sample size 50). We can see that the jackknife method produces the longest 95% CI followed by MI using Beta distribution approximation except when true success rate is extremely high, 99%. In terms of 95% coverage, complete case analysis, single imputation as all success, and our 3 proposed methods were able to produce 95% CI that covered the truth more than 95% of the time. When sample size increases with a 10% missing rate and 99% success rate, the average 95% CI length decreases for all methods. In general, complete case analysis, single imputation as all success, and our 3 proposed methods obtained nominal 95% coverage or slightly more. For the fully Bayesian method, 95% coverage was less than nominal when sample size is small, 10. Finally, as the missing rate increases, average 95% CI length increases for all methods, which is expected. Complete case, single imputation as all success, and our 3 proposed methods produced close to or more than 95% coverage under all the missing rate scenarios we investigated. 95% coverage for MI using Wald and MI using Wilson increased only when missing data was high, 30%. In summary, our simulation results suggest that our 3 proposed methods were robust and able to produce nominal 95% coverage and adequate or conservative average 95% CI length for all the simulation scenarios we investigated.

Figure 1: Average 95% confidence interval length and 95% coverage by true success rate under sample size of 50 and 10% missing rate.

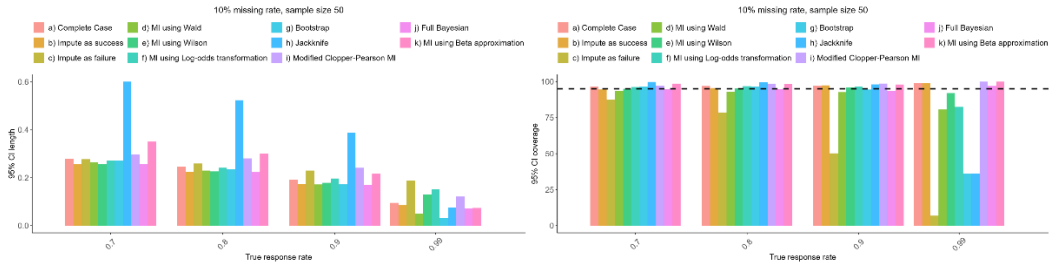

Figure 2: Average 95% confidence interval length and 95% coverage by sample size under 10% missing rate and 99% true success rate.

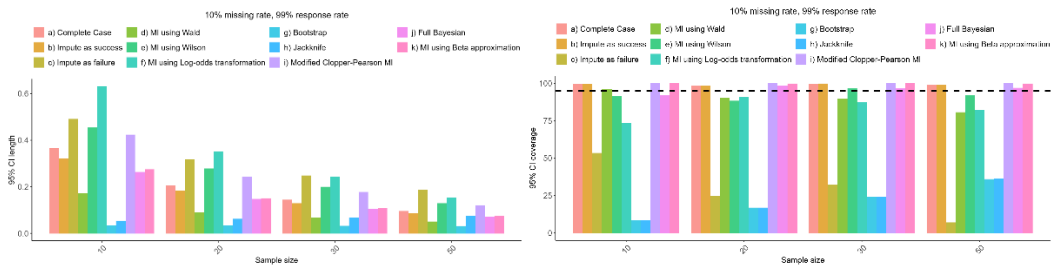

Figure 3: Average 95% confidence interval length and 95% coverage by missing rate under sample size 50 and 99% true success rate.

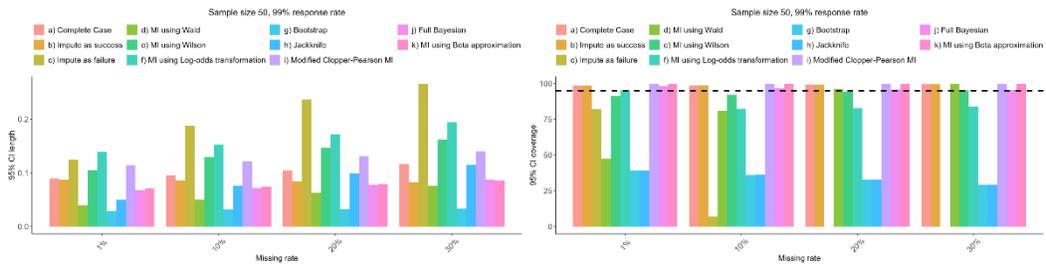

## 4. Data application

We now turn our attention to applying our proposed method together with the complete case analysis, single imputation as all success, single imputation as all failure, on a published clinical trial. The CIT Protocol 07 (CIT-07) is a prospective, multicenter, open-label, single-arm phase 3 study to evaluate the effectiveness and safety of the investigational product purified human pancreatic islets (PHPI) in adult subjects with Type 1 Diabetes (T1D) > 5 years and persistent impaired awareness of hypoglycemia (IAH) and severe hypoglycemic events (SHEs) despite medical treatment. It was conducted at eight centers in North America by the CIT Consortium (Hering et al., 2016). Forty-eight subjects received one or more transplants of PHPI.

The primary endpoint is the achievement of an HbA1c < 7.0% at Day 365 and freedom of SHEs from Day 28 to Day 365 inclusive following the first islet transplant with the day of transplant designated Day 0. The primary endpoint is assessed at Day 365 (Year 1) and another similar secondary endpoint is assessed at Day 730 (Year 2) with HbA1c < 7.0% at Day 730 and freedom of SHEs from Day 28 to Day 730. Forty-two of 48 subjects achieved the primary endpoint, 3 subjects failed to achieve the primary endpoint, and another 3 subjects had missing data due to lost to follow-up or withdrew from the study. At Year 2, 34 out of 48 subjects achieved the endpoint, 8 subjects failed, and 6 subjects had missing outcome due to lost to follow-up, early withdrawal and other administrative reasons.

4.1 Results

Figure 4 shows the forest plot of the 95% CI together with the estimate for the proportion of subjects achieving CIT-07 primary endpoint at Year 1 based on 1) the complete case analysis, 2) single imputation imputing all missing data as failures, 3) single imputation imputing all missing data as success, 4) standard MI using Wald intervals, 5) our modified CPMI approach, 6) MI using Beta approximation, and 7) the full Bayesian approach. Note that the 95% CI of 1) to 3) are based on Clopper-Pearson confidence interval with no or single imputation. All imputation methods produced fairly similar estimates (93%), except for single imputation as all failures (87.5%; note that Hering et al. [2016] imputed all missing data as failures). Hering et al. (2016) reported an estimate of 87.5% and a lower bound, one-sided 95% CI of 76.8%. Based on our simulation results, we recommend interpretation from the full Bayesian approach (93.0% [83.2%, 98.1%]), which often provide 95% CI coverage close to the nominal 95%. If conservative and valid results are desired, we recommend results from MI using Beta distribution (92.5% [81.6%, >99.0%]) based on our simulation results. Finally, we note that MI using Wald and modified CPMI produced 95% CI upper bounds more than 100% as the result of normal approximation. If intervals within (0%, 100%) are desired, Wilson-intervals, the fully Bayesian, or MI using Beta approximation will be able to produce such intervals. Figure 5 shows the forest plot of the 95% CI together with the estimate for the proportion of subjects achieving primary endpoint at Year 2. Hering et al. (2016) reported 71% without any 95% CI provided. Our full Bayesian approach and MI using Beta approximation reported 80.7% (67.4%, 90.6%) and 80.7% (64.7%, 95.0%) respectively.

Figure 4: Forest plot of 95% Confidence Intervals for proportion of subjects achieving CIT-07 primary endpoint at year-1 post-initial transplant under various missing data handling methods.

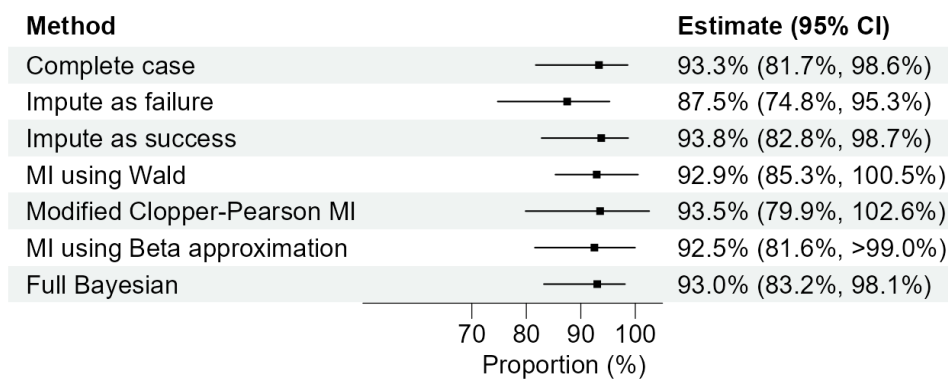

Figure 5: Forest plot of 95% Confidence Intervals for proportion of subjects achieving CIT-07 secondary endpoint at year-2 post-initial transplant under various missing data handling methods.

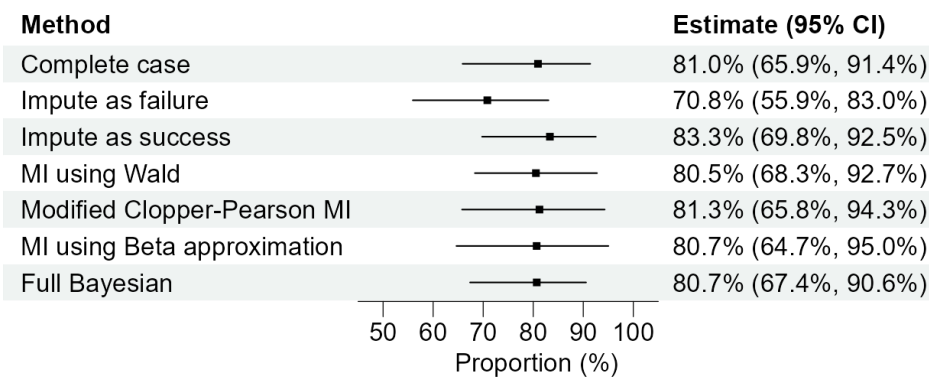

## 5. Conclusion

Small clinical trials with extremely high success rates were exceedingly rare until the recent emergence of CGT trials. Although sponsors try their best, missing data can still occur. As CGT methods are usually innovative, out of an abundance of caution, regulators will usually request sponsors to propose a strategy for handling missing data. Often, the MI approach is accepted due to its general applicability as well as proper accounting of missingness uncertainty. Unfortunately, literature regarding the proper implementation of MI to binary endpoints in situations of small sample size and extremely high success rates are scarce. In this work, we investigated various established missing data methods as well as proposed three methods for implementing MI on a binary endpoint under small sample size with extremely high success rates. We found that the fully Bayesian imputation produced the most efficient 95% CI estimate while maintaining proper 95% coverage except when sample size is small, 10. Approximating MI using Beta distribution and the modified CPMI approach both produced conservative 95% CI which had coverage more than 95% for nearly all the simulation scenarios we considered.

Applying our three proposed methods to a published clinical trial, CIT-07, and comparing with results from complete case analysis, single imputation as failures, single imputation as success, and standard MI using Wald intervals, we suggest to use results from our full Bayesian approach or MI using Beta distribution.

Throughout this work, we focused on the single arm trial design. As CGT matures, regulators may increasingly request sponsors to conduct a randomized control trial (RCT) instead. In RCT, although endpoints may still be binary, the parameter of interest is now the risk difference, risk ratio, or odds ratio. How to extend to such parameters will be an important future work. Another area of future work is when different patterns of missingness are needed for different subgroups of the study population for example, pattern mixture models using MI, retrieved dropout, or jump to reference methods. Using an example to elaborate, the reason for study withdrawal could be used to stratify the study population into various subgroups with each subgroup having its own assumption on the prior and likelihood. Whether the eventual joint posterior predictive distribution will still be a Beta-Binomial distribution, the joint posterior distribution draw will still be a Beta distribution, or whether the Beta distribution can still be used to approximate the MI will require further investigation. Finally, investigating whether the Beta-distribution assumption can be as general as the use of normal approximations for MI for parameters of interest with range of 0% to 100% will be another exciting area for future research.

Appendix

Complete case analysis – all subjects in sample with observed endpoint. Clopper-Pearson method was used to obtain the exact 95% CI.

Impute all missing data as success – subjects with missing endpoint imputed as success. Clopper-Pearson method applied to the observed and imputed data to obtain the exact 95% CI.

Impute all missing data as failure – similar to impute all missing data as success but instead impute as failures.

Multiple imputation using Wald intervals – draw $\pi$ and $y^{miss}$ using Equation (7). Use $E[\pi|Y^{obs}] \pm z_{1-\frac{\alpha}{2}}\sqrt{Var[\pi|Y^{obs}]}$ to obtain the 95% CI. The variance of each multiply impute dataset parameter is calculated as $V_d = \frac{\frac{y^{obs}+y^{miss}}{n}(1-\frac{y^{obs}+y^{miss}}{n})}{n}$.

Multiple imputation using Wilson intervals – refer to Lott and Reiter (2020) for implementation details.

Multiple imputation using log-odds transformation with Barnard and Rubin (1999) adjustment – draw $y^{miss}$ using Equation (7). Perform a log-odds transformation to obtain $\theta^d = \log\left(\frac{\frac{y^{obs}+y^{miss}}{n}}{1-\frac{y^{obs}+y^{miss}}{n}}\right)$. Perform MI on $\theta^d$ with $V_d$ estimated as $V_d = \frac{1}{y^{obs}+y^{miss}} + \frac{1}{n-y^{obs}-y^{miss}}$. Use $t_{v^*}$ where $t$ is the $t$-distribution adjustment for small $D$ and $v^*$ is the Barnard and Rubin (1999) small sample degrees of freedom adjustment.

Bootstrap – Resample data with replacement, $B=1000$ times to create 1000 $n$ sample size data. Run analysis on these 1000 datasets with $\hat{\pi}^b = \frac{y^{b,obs}+y^{b,miss}}{n}$, where $y^{b,miss}$ is the imputed number of success by drawing from a binomial distribution with $Bin(n^{b,miss}, \frac{y^{b,obs}}{n^{b,obs}})$ at the $b^{th}$ bootstrap. The 95% CI is given by $I_{emp}(\pi) = (\hat{\pi}^{(b,\frac{\alpha}{2})}, \hat{\pi}^{(b,1-\frac{\alpha}{2})})$.

Jackknife – Let $\hat{\pi}^{/j} = \frac{y^{j,obs}+y^{j,miss}}{n}$ be the $j^{th}$ jackknife calculated by removing the $j^{th}$ subject. The jackknife estimate is calculated as $\hat{\pi} = \frac{1}{n}\sum_{j=1}^{n}\hat{\pi}^{/j}$. The jackknife 95% CI is calculated as $\hat{\pi} \pm z_{1-\frac{\alpha}{2}}\sqrt{Var(\hat{\pi})}$ where $Var(\hat{\pi}) = \frac{n-1}{n}\sum_{i=1}^{n}(\hat{\pi}^{/j} - \hat{\pi})^2$.

Modified Clopper-Pearson multiple imputation – $D = DD = 50$ was used to obtain the modified CPMI.

Full Bayesian imputation – $M = 5000$ was used to obtain the full Bayesian draw due to computation limitations.

Multiple imputation using Beta distribution – $D = 50$ was used to compute the MI and 0.001 increments were used to conduct the grid search to obtain the 95% CI .

5000 simulations were conducted for each scenario.